\documentstyle[epsfig,12pt]{article}
\def\thefootnote{\fnsymbol{footnote}}
\begin{document}
\begin{titlepage}
\today          \hfill 
\begin{center}
\hfill    LBNL-48165 \\
\hfill    UCB-PTH-01/16 \\
\hfill hep-th/0105098 \\

\vskip .5in
\renewcommand{\thefootnote}{\fnsymbol{footnote}}
{\Large \bf Tachyon condensation in boundary string field theory at one
loop} \footnote{This work was supported by the Director, 
Office of Energy 
Research, Office of High Energy and Nuclear Physics, Division of High 
Energy Physics of the U.S. Department of Energy under Contract 
DE-AC03-76SF00098 and in part by the National Science Foundation grant PHY-95-14797.}
\vskip .50in

\vskip .3in
 Korkut Bardakci\footnote{email address: kbardakci@lbl.gov} and 
Anatoly Konechny\footnote{email address: konechny@thsrv.lbl.gov}

\vskip 0.5cm
{\em Department of Physics\\
University of California at Berkeley\\
   and\\
 Theoretical Physics Group\\
    Lawrence Berkeley National Laboratory\\
      University of California\\
    Berkeley, California 94720}
\end{center}

\vskip .2in

\begin{abstract}

We compute the one-loop partition function for quadratic tachyon background in open string theory.  
 Both closed and open string representations are developed. 
Using these representations  we study the one-loop divergences in the partition function 
in the presence of the  tachyon background. The divergences due to the open and closed string 
tachyons are treated by analytic continuation in the tachyon mass squared. We pay particular 
attention to the imaginary part of the analytically continued expressions. The last one 
gives  the decay rate of the unstable vacuum. The  dilaton tadpole is also 
given some partial consideration.
The partition function is further used to  study   corrections to tachyon condensation processes 
describing  brane descent relations. Assuming  the boundary string field theory prescription 
for construction of the string field action via partition function holds at one loop level 
we study the  
  one-loop corrections to the tachyon potential and to the tensions of lower-dimensional branes. 
\end{abstract}

\end{titlepage}

\newpage
\renewcommand{\thepage}{\arabic{page}}
\setcounter{page}{1}

\section{Introduction}
First attempts to find a stable nonperturbative vacuum in bosonic string theory were 
made way back in the 70's \cite{Bardakci1},  \cite{BardHalp}, \cite{Bardakci2}. 
Part of the complexity of the problem comes from the fact that tachyon condensation is 
an off-shell problem. With the advent of string field theory new attempts were made 
\cite{Sam1}, \cite{Sam2} that studied tachyon potential and 
provided more evidence that a new stable vacuum indeed 
exists. Later an important insight into the problem was provided by A. Sen \cite{Sen} 
who among other things put forward a conjecture that the hight of the tachyon potential 
in open bosonic string theory 
is equal to the tension of the space-filling  D25-brane. This point of view as well as string field theory methods 
were further developed 
in a series of papers 
\cite{Sen-1}, \cite{Sen1}, \cite{Sen2}, \cite{SenZw}, \cite{MTaylor} (and  
references therein). 
The tachyon condensation 
is believed  to yield a complete decoupling of the open string states.
 The D25-brane desintegration into the closed string vacuum  may go through various 
metastable phases described by lower-dimensional branes \cite{Sen1}. These descent relations are 
in general easier to study than the complete condensation. 

In the papers  cited above the cubic string field theory was one of the primary tools of investigation. 
Recently it was realized that another version of open string theory nowadays christened as Boundary 
String Field Theory (BSFT) can be very useful in studying the question. The BSFT was put forward 
by E.~Witten in \cite{Witten1} and further developed in the papers of   
E.~Witten \cite{Witten2},  K.~Li and E.~Witten \cite{Witten3}, and S.~Shatashvili \cite{Shat1}, 
\cite{Shat2}. Using BSFT methods the exact tree level tachyon potential was derived  in 
\cite{Gerasimov1}, \cite{Kutasov1} and the Sen's conjecture regarding the hight of the potential was shown 
to be true. In particular BSFT was shown to describe most elegantly  brane descent relations. 
We would like to  note that the BSFT in its spirit is very similar to the old sigma model approach 
(see  \cite{Tseytlin1}, \cite{Tseytlin2} for a review and  \cite{Tseytlin3} for a recent discussion). 
The picture of tachyon  condensation in bosonic   BSFT  (as well as the 
BSFT itself) was further developed in \cite{Gerasimov2}, \cite{Gosh_Sen}, \cite{Shenker}, 
\cite{Andreev1}, \cite{Andreev2}, \cite{Frolov}.

In this paper, we investigate one loop corrections to the effective 
action of the tachyon field, probed by the mixed boundary
conditions
\begin{equation} \label{quadratic}
\frac{\partial X}{\partial\tau}=u X,
\end{equation}
first studied  in \cite{Witten2}  and later used in many subsequent papers.
 There are several motivations for
studying this problem; for example, one would like to see whether the 
system stays weakly coupled as the tachyon rolls down the potential
and also one would like to test Sen's conjectures. Our aim in this paper
is more modest; we wish to carry out a divergence free and internally
consistent one loop correction to the tachyon potential and the D-brane
tension.
In our analysis we take the approach of 
BSFT.
BSFT gives a (background independent)  prescription of how to compute a space-time 
action in the presence of an open string (off-shell) background. The prescription was only developed 
at  tree level string theory. In view of the lack of a general theoretical foundation of a quantum BSFT, we 
proceed with a speculative procedure for computing the effective space-time action that extends 
the tree level prescription in the most direct manner. The key ingredient in the computation is 
 a one-loop partition  function in the presence of tachyon background (see section 2 for a discussion). 
 This   amplitude can be considered as a closed string
propagating at tree level for a (Euclidean) time T
 between initial and final states representing the
boundary conditions (boundary states). This is represented by a
cylinder graph. We choose the boundary conditions
to be independent of the time T in this picture. In section 3, the cylinder
is mapped into an annulus by a conformal tranformation, and the 
resulting boundary conditions are shown to depend on T. We should point
out that the imposition of the simple (T independent) boundary
conditions in the cylinder picture, as opposed to, say, in the
annulus picture, is somewhat  arbitrary, although in our opinion, a natural
choice. In section 4, we compute the cylinder amplitude up to an overall
normalization constant using BSFT, and in section 5,
 we perform an independent check on our result by computing the same
amplitude in the annular region.

 We should stress that the calculations described so far were carried
out at a fixed modulus, which is the time T for the cylinder or the
ratio of the two radii for the annulus. 
To complete the
calculation of the one loop partition function, we have to add the
contribution of the ghost sector and integrate over the modulus
with a suitable measure. In the case of the usual boundary conditions
(Neumann or Dirichlet), conformal invariance uniquely determines
both the ghost contribution and the modular measure. Since in our
problem the boundary conditions violate conformal invariance, we 
know no convincing way of uniquely fixing these contributions. In
the absence of a guiding principle, we have decided to keep the
ghost contribution the same as in the conformal case, and use the
same conformal measure in the integration over the modulus T in
the cylinder picture (see eq. (\ref{part_f})). Although this is an ad hoc 
recipe, it has the virtue of being simple and having the correct
limit as $u\rightarrow 0$ (Neumann boundary). We should point out
that in principle there is an ambiguity
 even in the calculation of the tree level open string
amplitude with non-conformal boundary conditions [13]. The result
of the calculation would in general depend on the region of the world
sheet chosen; for example, a calculation done using the upper half plane
would give a different result than the one done using a circle. This is a 
consequence of the lack of invariance under the conformal tranformation
connecting the two regions. Of course, this does not mean that
the ambiguity in the calculation of the amplitude corresponds to an
ambiguity in the resulting physics. Since the calculation of the tree
amplitude in [13] rests on firm foundation, namely BRST invariance [12],
it is generally believed that amplitudes calculated in different
regions must be related by field redefinitions, which do not change
the underlying physics. However, in the case of the one loop
amplitude, there is no such well founded starting point, and 
our naive prescription may need to be modified in the future\footnote{ After this 
work was completed preprints \cite{Larsen}, \cite{Semenof} appeared that propose a different scheme for 
computing loop corrections in BSFT.}. 
 In spite of these
reservations, we believe that it is of some interest to carry  the
calculation to the end to find the correction to the tachyon
potential. As we shall argue later, the final results appear to be
reasonable and self consistent.

An alternative way of looking at the cylinder amplitude is to view it as the 
calculation of the partition function of an open string with mixed
boundary conditions corresponding to (\ref{quadratic}). 
This calculation, which is technically more involved
then the calculation of the cylinder amplitude, is carried out in
section 8. The
final answer is in a partially implicit form difficult to compare with the 
earlier result in detail. However, the open string picture has its
 advantages. For
example, some of the divergences of the amplitude are easier to
handle, and the undetermined overall constant of the previous
calculation is easily fixed.
 
We would like to remark that the computation 
of the boundary state and the one-loop partition function in two channels is 
essentially independent of its further use in the construction of space-time 
effective action and we believe it to be of interest by itself.

Finally, we would like to discuss briefly the divergences encountered
in the integration over the modulus. These divergences are caused by
the tachyons present in both the open and closed string channels, and
by the dilaton in the closed string channel. We think of the
divergences due to the tachyons as being similar to the superficial
divergences encountered in the integral representation of the tree
level string amplitudes. These latter divergences are easily circumvented
by appropriate analytic continuations in external momenta.
 The same idea of analytic continuation can be used for the tachyonic
divergences [27], with a resulting complex tachyon potential. This is, of
course, due to the instability of the vacuum in the presence of
the tachyon. An alternative approach, which we will not use, is to
cancel the tachyon divergence by a tree level counter term (Fischler-
Susskind mechanism \cite{FS}). The divergence due to the dilaton, however, has
to be canceled by the Fischler-Susskind mechanism
 when it is present.  However, in this paper we will restrict ourselves 
to the situation when  there is no dilatonic divergence and no need for tree level
counter terms. This happens in the process describing the descent relation of D25 brane 
to a D25-p brane with $p>2$.

The paper is organized as follows. In section 2 we give a general discussion of BSFT and the loop corrections in it. 
Section 3 contains a further discussion of the boundary conditions at one loop corresponding to the quadratic 
tachyon perturbation. In section 4 we compute the corresponding boundary state and find the expression for  
partition function in the closed string channel. In section 5 we give an alternative computation via Green's 
function on the annulus and discuss renormalization conditions. In section 6 we remind the reader about the situation 
with one-loop divergences in the conformal case and about the analytic continuation treatment of the tachyonic divergences.
The divergences in the closed string channel in the presence of the tachyon background are considered  in section 7. 
In section 8 we develop the open string channel description  of the partition function. 
We approximately compute the modified Casimir energy for the  boundary conditions describing the open string channel. 
The open string tachyonic divergence  comes from the contributions of the ground state (whose energy is given by the Casimir 
energy) and the first excited state. We derive a general integral representation for the Casimir energy as well 
as an approximate expression valid for small coupling constant. These results are used then to study the   
general form of the  tachyon divergence.  
In section 9 we discuss 
the dilaton divergence and compute the one-loop correction to the tachyon potential. In section 10 we derive 
the tensions of the lower-dimensional branes by finding the asymptotic value of the action in the limit  
$a, u\to \infty$. We conclude with a discussion and a list of unsolved questions in section 11.    


\section{Bosonic boundary string field theory} 
The starting point of the original paper \cite{Witten1} in which BSFT were introduced was a  
Batalin-Vilkovisky (BV) formalism on the space of sigma model boundary perturbations. Consider 
a world sheet action defined on a unit disc on the complex plane with standard metric that  has   the form 
\begin{equation}\label{sigma}
S = S_{0} + \int_{\partial \Sigma} d\phi \, {\cal V} 
\end{equation}
that is $S$ is equal to a sum of the standard free action $S_{0}$ in the bulk 
corresponding to a fixed closed string background and a boundary perturbation 
specified by some local operator of ghost number zero ${\cal V} $ constructed 
from  the fields $X_{\mu}$ and ghosts $b, c$. 
The space of such operators is considered to be a phase space of the BSFT. Note that the lack of precise definition 
of the admissible class of operators ${\cal V} $ (or equivalently a space of  boundary conditions) 
is still a major problem of BSFT.

A further assumption is that  ${\cal V} $ can be represented as ${\cal V} = b_{-1}{\cal O}$.
In the situation when ghosts and matter are decoupled one has 
${\cal O} = c{\cal V}$. 
(Strictly speaking in Witten's formulation $\cal O$ is the main object specifying a point in the phase space. 
However we will soon assume that matter and ghosts decouple and work only with $\cal V$'s.) 
Then the Witten's BV antibracket is defined as 
$$
\omega ( \delta {\cal O}_{1}, \delta {\cal O}_{2} ) = \int d\phi_{1} \int d\phi_{2} 
\langle \delta {\cal O}_{1}(\phi_{1})  \delta {\cal O}_{2}(\phi_{2}) \rangle    
$$ 
where $\langle ... \rangle$ stands for a correlator in the presence of the background  $\cal V$ ($\cal O$), i.e. 
the point in phase space at which we evaluate $\omega$. 

The string field action $\cal S$  is defined as a Hamiltonian  for the vector field 
specified on the phase space by the BRST operator $Q$ that is assumed to be determined by the standard 
bulk part $S_{0}$. Let us expand the boundary perturbation $\cal O$ in terms of  some basis ${\cal O}_{i}$: 
$$
{\cal O} = \sum_{i}\lambda^{i}{\cal O}_{i}
$$
where $\lambda^{i}$ are coupling constants (coordinates on the phase space). 
We can write then the following equation for   $\cal S$  
\begin{equation} \label{action1}
\frac{\partial {\cal S}}{\partial \lambda^{i}}  = \frac{1}{2} \int d\phi_{1}\int d\phi_{2} \, 
\langle {\cal O}_{i}(\phi_{1}) \{ Q, {\cal O}(\phi_{2}) \} \rangle \, . 
\end{equation}
Since $Q^{2}=0$ the string field  action $\cal S$ should satisfy the (classical) master equation
$$
\{ {\cal S}, {\cal S} \}_{BV} = 0 
$$
where $\{ . , . \}_{BV}$ is the BV bracket specified by $\omega$.  

The action $\cal S$ is a string field tree-level  action. A natural way to extend Witten's formulation 
to the full quantum theory would be to consider a quantum master equation 
$$
\hbar \Delta_{\rho} {\cal S} + \frac{1}{2}\{ {\cal S}, {\cal S} \}_{BV} = 0 \, . 
$$
To specify the operator  $\Delta_{\rho}$ one needs to supply the phase space with a measure 
$\rho$ \cite{Schwarz}. 
Then  $\Delta_{\rho}$ is an operator that acting on a function $A$ gives the divergence 
$$
{\rm div}_{\rho} A =  \frac{1}{\rho}\partial_{\lambda_{i}}(\rho A^{i}) 
$$ 
where $A^{i}$ is a vector field corresponding to $A$. The loop expansion then would correspond 
to the expansion of $\cal S$ in powers of $\hbar$. 
The density $\rho$ is essentially an independent ingredient in the formalism (see \cite{Zwiebach1} for 
a thorough discussion). Lacking a rigorous  definition of the phase space itself the idea of finding some 
natural measure on it does not look very promising at the moment. Thus we will have to proceed in some other way 
to define the loop corrections. 

From now on we will talk only about the situation when matter and ghosts are decoupled and ${\cal O} = c{\cal V}$. 
In paper \cite{Witten2} (see also \cite{Shat1}) 
it was shown on general grounds that a solution to (\ref{action1}) (for the case when 
matter and ghosts are decoupled) must be of the form 
\begin{equation} \label{S1}
{\cal S} = (1 + V^{i}\frac{\partial}{\partial \lambda^{i}}) Z
\end{equation}
where $Z$ is the disc partition function corresponding to (\ref{sigma}) and $V^{i}$ is some vector field on the 
space of open string fields $\cal V$. The equation of motions following from this action are always linear. 

It was shown later by Shatashvili  \cite{Shat2} that a more careful treatment 
of total derivatives in correlation functions leads to a natural modification of (\ref{S1}) allowing 
nonlinearities in the equations of motion and proposed the 
following relation  
\begin{equation} \label{S2}
{\cal S} = (1 + \beta^{i}\frac{\partial}{\partial \lambda^{i}}) Z
\end{equation}
where $\beta^{i}$ is the beta function corresponding to coupling constant $\lambda^{i}$. This relation was 
shown to be true in the first order in conformal perturbation theory  \cite{Shat2}. 
Note that in order to account for the nonlinear contributions in $\beta_{i}$ (contributions of contact terms) 
in the framework of the original BV formalism one has to modify the BRST operator $Q$ that now has to be 
dependent on $\lambda^{i}$. 
As long as the beta function is linear (and one can always choose locally such set of coordinates $\lambda^{i}$ 
when this is true) the equations (\ref{S1}) and (\ref{S2}) seem to be equivalent. However that describes only 
one coordinate patch in the whole phase space manifold. In particular the set of coordinates in which the 
beta function is linear is singular when  perturbations $\cal V$ approach the mass shell. But as long as we stay far 
off shell that is a situation of primary interest in the case with tachyon condensation this coordinate 
system  works well. (See \cite{Kutasov1}, \cite{Frolov} for a discussion of the on shell behavior of  BSFT.)

We will take  formula (\ref{S2}) as a starting point for constructing the BSFT action. 
Written in that form the BSFT  action can be easily linked with the sigma model approach 
(see \cite{Tseytlin1} for a general review and \cite{Tseytlin3} for a recent discussion on the relation 
of the sigma model approach and BSFT). Indeed it was noticed a long time ago \cite{Fradkin_Ts}
that being a generating functional for scattering  amplitudes the renormalized sigma model partition function is a 
natural candidate for string theory effective    action. And this identification works quite well in the vicinity 
of the mass shell of massless particles. However if one wants to include tachyons in the sigma model approach (that is 
natural because perturbations corresponding to tachyons are relevant and do not change the renormalizability 
unlike the massive string fields) than the identification ${\cal S} = Z^{ren}$ does not work. It does not 
give the correct equations of motion because 
$$
\frac{\partial Z}{\partial \lambda}  = \int d\phi \langle {\cal V}(\phi) \rangle 
$$  
does not vanish in general at the conformal point $\lambda=0$ if $\cal V$ has conformal dimension 1 that 
is for example the case for the constant tachyon mode. 
The second term in the expression (\ref{S2}) corrects that problem. Thus if we substitute 
$ \beta({\cal \lambda}) = -\lambda + {\cal O}(\lambda^{2})$ we will get 
$$
\frac{\partial {\cal S}}{\partial \lambda}   =  
\left( \frac{\partial}{\partial \lambda}{\cal O}(\lambda^{2}) \right)Z
$$   
that evidently vanishes at the original fixed point $\lambda = 0$. This means that in a coordinate patch in which the 
beta function is linear (\ref{S2}) gives the correct equations of motion. It is believed that in general there 
exists a nonsingular metric $G_{ij}$ defined on the whole manifold of string fields such that 
$$
\frac{\partial {\cal S} }{\partial \lambda^{i}}  = G_{ij}(\lambda) \beta^{j} \, . 
$$

In the sigma model approach a generating functional for scattering amplitudes that includes all string loop 
corrections is given by  the total renormalized sigma model partition function. For the open string theory it 
has the form  
$$
Z = \sum_{b} \frac{g^{ -1 + b}}{b!} Z_{ b} 
$$ 
where the sum is over world sheets with  $b$-boundaries. 
Moreover the beta functions of massless fields are known to receive loop corrections coming from modular infinities (see 
\cite{Tseytlin2} for a review). We see then that  from the sigma model point of view formula (\ref{S2}) 
has a natural generalization that includes loop corrections. 

It should be noted that an off shell extension of the sigma model approach to string theory involves a 
great deal of arbitrariness having to do with gauge fixing and field redefinitions. When doing loop 
corrections the world sheet metrics and the sigma model backgrounds need to be chosen consistently at each order 
of perturbation theory. Since in the problem at hand the bulk part of the sigma model action is fixed 
(that corresponds to a fixed closed string background) it seems to be natural to integrate over the moduli 
using a closed string picture of the amplitude at each order. In this picture we consider a $b-1$-loop 
open string vacuum amplitude as a $b$-point tree level scattering of closed string states $|{\cal V}\rangle$ 
specified by the open string background functional ${\cal V}$. In the case when the perturbation ${\cal V}$ is
conformal this correspondence is well established and the corresponding closed string state $|{\cal V}\rangle$ is 
called a boundary state \cite{Callan1}. The normalization of  $|{\cal V}\rangle$ is fixed by the 
equality of open and closed string channel representations for the one-loop partition function 
\begin{equation}\label{bs1}
 {\rm Tr} e^{-H^{open}_{\cal V}T} = \langle {\cal V}|e^{-H^{cl.}_{0}\pi/T} |{\cal V}\rangle \, . 
\end{equation}
In the left hand side of this equation we have an open string partition function 
 on a strip of length $T$ with periodic boundary conditions in the time direction and 
the perturbed open string Hamiltonian $H^{open}_{\cal V}$. In the right hand side  we have a closed string  amplitude 
 evaluated  on a cylinder of length $\pi/T$ using the free 
closed string Hamiltonian $H^{cl.}_{0}$. The overlap of  $|{\cal V}\rangle$ with the closed string 
Fock space vacuum $|0\rangle$ is called a boundary entropy \cite{AL} and is proportional to the 
disc partition function 
\begin{equation}\label{bs2}
\langle 0|{\cal V}\rangle \sim Z_{disc} ({\cal V}) \, . 
\end{equation}
In this paper we construct a boundary state satisfying (\ref{bs1}), (\ref{bs2}) for the case of quadratic 
tachyon perturbation on the boundary. Lacking a general quantum BSFT theory we  explore an  ad hoc 
 prescription for a  one loop corrected BSFT effective space-time action  
that uses  formula (\ref{S2}) and the one loop partition function. Note that this construction is guaranteed to 
give the correct one loop corrections to the on-shell amplitudes. However apriory it is not clear whether 
this prescription has all of the desired properties for the truly off-shell quantities. 
We believe that whatever the correct  one loop prescription      
may be many of the  features present in our speculative construction, such as the imaginary part of the tachyon potential and 
loop corrections to the brane tensions, should remain.  
In the next section we will discuss in more detail 
the 1-loop boundary conditions corresponding to the quadratic tachyon background.


\section{Boundary condition} 
As it was discussed in \cite{Kutasov1} a boundary perturbation corresponding to quadratic 
tachyon profile is particularly useful  for  describing descent relation between unstable D-branes. 
The quadratic profile has a unique property of preserving its shape along the RG flow, i.e. the corresponding  
modes (coupling constants) can be consistently decoupled from all other string modes. 
  
In the conventions of Witten's paper \cite{Witten2} the particular background we are interested in 
is specified (at tree level) by  the following action on a unit disc  
\begin{equation} \label{tree_action}
S = \frac{1}{8\pi } \Bigl( \int d\phi \int r dr \partial_{\alpha} X \partial^{\alpha} X   
+ u \int_{0}^{2\pi }d\phi X^{2} (\phi )\Bigr) + a  
\end{equation}
which is written  for a single string coordinate field $X$.
Here $r, \phi$ are polar coordinates, $u$ and $a$ are (nonnegative) constants. 

Due to the lack of conformal invariance it is not immediately obvious what boundary conditions describe the same 
background at the loop level. The disc representation (\ref{tree_action}) has to do with some particular off-shell 
gauge fixing. Since the boundary term in  (\ref{tree_action}) is rotationally invariant it is 
naturally to expect that at the one-loop level the same background must be represented 
by boundary conditions on an annulus $ r_{0}\le |z| \le 1$, $z=re^{i\phi}$ in such a way
 that on each circle $|z| = 1$ and $|z| = r_{0}$ 
the boundary term is equivalent to the one in (\ref{tree_action}). A conformal transformation that interchanges the 
two circles is the inversion $z \mapsto r^{2}_{0}/ \bar z$. So one can take  
for  the circle $|z| = 1$ exactly the same boundary term as at the tree level  
and on the circle  $|z| = r_{0}$ the one obtained from it by the aforementioned conformal transformation. 
The boundary conditions then read
\begin{equation} \label{bc}
\frac{\partial}{\partial r} X + u X  = 0\, , \enspace r=1 \, , \qquad   
-r_{0}\frac{\partial}{\partial r} X + u X  = 0 \, , \enspace  r=r_{0} \, .  
\end{equation}
The one-loop partition function for the quadratic tachyon potential was considered in a number of papers
\cite{Vish1}, \cite{Su}, \cite{Vish2}. 
The boundary conditions on the annulus considered in papers \cite{Vish2}, \cite{Alish} (section 4)
 match with ours. The other papers consider boundary conditions that either 
 do not have a relative sign in (\ref{bc}) that 
we think is quite unnatural from the point of view that tachyon does not carry any charge, or do not have 
a factor of $r_{0}$ in the second condition in which case we think the two boundaries are not treated on 
equal footing. 

 These  boundary conditions take the most transparent form on  a cylinder  with coordinates 
$ \sigma = \phi $, $\tau = {\rm ln} r $ with ranges $0 \le \sigma < 2\pi$, $ - t \le \tau \le 0$, 
$t = -  {\rm ln} r_{0}$.  
They can be represented by means of a boundary state $|u\rangle$  in the Fock space of the first quantized 
closed string theory. It is defined up to an overall  constant by the  equation
\begin{equation} \label{bc2}
\frac{\partial X}{\partial \tau} |u\rangle_{\tau} =  u X|u \rangle_{\tau} \, .
\end{equation}
Then the boundary condition (\ref{bc}) is represented on the cylinder by the boundary state $|u\rangle_{\tau = 0}$ 
being the initial state at the right end of the cylinder and by the conjugated state $\langle u|_{\tau = -t}$ 
being the final state at the left end. The  partition function for the cylinder of length $t$ is then given by the expression 
\begin{equation} \label{1loop_t}
Z_{1}(u, a, t) = e^{-2a}  \langle u| e^{(L_{0} + \tilde L_{0})t} | u \rangle    
\end{equation}
where $|u\rangle \equiv |u\rangle_{\tau = 0}$. 

Note that the tree level partition function $Z_{0}(u, a)$ corresponding to  
the action  (\ref{tree_action}) is given up to an overall  numerical constant $C$ to be discussed later 
by the overlap of $|u\rangle$ with the closed string vacuum  
\begin{equation} \label{b_entr}
Z_{0}(a, u) = C\cdot e^{-a} \langle 0|u\rangle 
\end{equation}
where we prefer not to include the factor $e^{-a}$ in the definition of $|u\rangle$. The overlap itself 
in the conformal situation gives the value of the boundary entropy \cite{AL}.
Equation (\ref{b_entr}) along with   equation (\ref{bc2}) allows one to compute $|u\rangle$ up 
to an overall numerical constant. 
The inversion used on the annulus corresponds to the reflection about the middle of the cylinder that interchanges 
the two ends of the cylinder. Evidently the boundary conditions are symmetric with respect to this reflection.

It is instructive to note that the tree level representation as it is fixed on the unit disc favors the cylindrical 
quantization. 
 For instance the same boundary conditions on an infinite strip will be time dependent.   
This suggests that quite generally in BSFT one may think 
of the open string configuration space as some suitable space of boundary states.

Consider now a conformally equivalent cylinder with coordinates 
$\sigma' =  - \tau \cdot \frac{\pi}{t}  $, $\tau' = \sigma\cdot \frac{\pi}{t} $ ranging as 
$0\le \sigma' \le \pi$, $0\le \tau' \le  \frac{2\pi^{2}}{t}\equiv 2\pi T$ with the boundaries $\tau' =0$ and 
$\tau' = 2\pi T$ identified.  

The boundary conditions (\ref{bc2}) being mapped on this cylinder take the form  
\begin{equation} \label{bc3}
-\frac{\partial X}{\partial \sigma'} + \frac{u}{T}X = 0\, , \enspace \sigma'=0 \, , \qquad 
\frac{\partial X}{\partial \sigma'} + \frac{u}{T}X = 0\, , \enspace \sigma' = \pi 
\end{equation}
where $T= \frac{\pi}{t}$. 

This picture corresponds to an open string with boundary conditions (\ref{bc3}) propagating in a loop of 
length $2\pi T$ in Euclidean time. The nonlocality in time of the boundary conditions (\ref{bc3}) stems 
from the fact that the boundary perturbation at hand  is not conformally invariant. Still it is relatively easy to 
quantize the theory in this representation that is quite useful in studying the partition function in the 
appropriate region of the moduli space.

The full one-loop partition function is obtained from (\ref{1loop_t}) by integration over the moduli space. We 
assume the measure of integration to be the same as in the conformal case, given by the appropriate ghost determinant. 
In general we will consider all $26$ coordinates of the string $X_{\mu}$ with the boundary condition being either 
the ones specified by the background (\ref{tree_action}) or the usual Neumann ones. 
In that case the sigma model action on the annulus has the form 
$$
S =  \frac{1}{8\pi } \Bigl( \int d\phi \int r dr \sum_{\mu = 1}^{26}\partial_{\alpha} X_{\mu} \partial^{\alpha} X^{\mu}   + 
\sum_{i=1}^{D} u_{i} ( \int_{|z|=1} d\phi X^{2}_{i} (\phi ) + \int_{|z|=r_{0}} d\phi X^{2}_{i} (\phi ) ) \Bigr) + 
2a \, .   
$$
The background is thus specified by the coupling constants $u_{i}$, $i=1, \dots , D$, $a$.
The full boundary state corresponding to these boundary conditions is then a tensor 
product 
$$
|B\rangle = \prod_{\mu =1}^{26}|B_{i}\rangle \, , \enspace |B_{i}\rangle = |u_{i}\rangle \, , i=1, \dots , D \, , 
 \enspace  
 |B_{\mu}\rangle = |N\rangle \, , \enspace \mu > D\, .  
$$
With this notation in mind we can write now the full one-loop partition function as 
\begin{equation} \label{part_f}
Z_{1}(u_{i}, a) = e^{-2a} 
\int_{-\infty }^{0} \frac{dt}{2\pi} e^{2t} f^{2}(-t)  \langle B| e^{(L_{0} + \tilde L_{0})t} | B \rangle 
\end{equation}
where 
$$
f^{2}(t) = (f(t))^{2} = \prod_{n=1}^{\infty} (1 - e^{-2tn})^{2} 
$$
comes from the ghost determinant.


\section{The partition function in the closed string channel}
We now proceed to calculate the boundary state (\ref{bc2}) and the partition function in the closed string channel. 
The mode expansion for a single coordinate field $X$ has the form 
\begin{equation}
X = \hat q - 2i\hat p\tau  + \sum_{1}^{\infty} \frac{1}{\sqrt{m}}\Bigl[ 
a^{\dagger}_{m} e^{m\tau + im\sigma} + \tilde a^{\dagger}_{m}e^{m\tau - im\sigma} + a_{m}e^{-m\tau - im\sigma} 
+ \tilde a_{m}e^{-m\tau + im\sigma} \Bigr] 
\end{equation}
and the commutation relations are 
$$
[\hat q, \hat p] = i \, , \qquad [a_{m}, a^{\dagger}_{n}] = \delta_{mn} \, . 
$$
Our conventions here coincide with the ones in \cite{Callan1} that correspond to the choice $\alpha'=2$ (in 
accord with the factor $1/8\pi$ in front of the action (\ref{tree_action})).
Substituting this mode expansion into (\ref{bc2}) we obtain 
\begin{eqnarray*}
i\hat p |u\rangle & = & \frac{-u}{2(1 - u\tau)}\hat q |u\rangle \, , \\
\tilde a_{m} |u\rangle & = &  \left( \frac{m-u}{m+u} \right) e^{2m\tau}a_{m}^{\dagger} |u\rangle \, , \\
 a_{m} |u\rangle & = &  \left( \frac{m-u}{m+u} \right) e^{2m\tau} \tilde a_{m}^{\dagger} |u\rangle \, . 
\end{eqnarray*} 
It is easy now to find a solution to these equations  
\begin{equation} \label{b_state}
|u\rangle_{\tau} = N(u) \cdot \frac{\sqrt{u}}{\sqrt{1- u\tau}} \cdot 
exp\Bigl( \sum_{m=1}^{\infty} \left( \frac{m-u}{m+u} \right) a_{m}^{\dagger} \tilde a_{m}^{\dagger} e^{2m\tau} \Bigr)
exp\Bigl(- \frac{u\hat q^{2}}{4(1-u\tau)} \Bigr) |0\rangle \, . 
\end{equation}
Here we inserted the factor $ \frac{\sqrt{u}}{\sqrt{1- u\tau}}$ 
corresponding to the normalization of the spreading Gaussian wave 
packet $exp(- \frac{uq^{2}}{4(1-u\tau)})$ 
that gives the correct time evolution.  The time independent constant $N(u)$ can be obtained up to a numerical factor 
from equation (\ref{b_entr}). The tree level partition factor for the boundary interaction (\ref{tree_action}) 
was found in \cite{Witten2} to be equal to 
\begin{equation}
Z_{0}(u) = \frac{1}{\sqrt{u}} \prod_{k=1}^{\infty} \frac{1}{1 + u/k}e^{u/k} = \sqrt{u}e^{\gamma u}\Gamma(u) 
\end{equation}
where $\gamma$ is the Euler constant. 
We keep the overall normalization of $Z_{0}(u)$ as it appeared in  \cite{Witten2}. 
Thus by (\ref{b_entr}) we have $N(u) = N_{0}\cdot Z_{0}(u)$ where $N_{0}$ is a numerical  constant. 
An analogous boundary state in the supersymmetric case was computed in \cite{deAlwis}. 

Plugging  the boundary state (\ref{b_state}) into formula  (\ref{part_f}) we obtain 
\begin{equation} \label{part_f2}
Z_{1}(u, a) = e^{-2a}\langle u|_{\tau= -t}u\rangle_{\tau=0} =   (N_{0})^{2} \cdot  (e^{-a}Z_{0}(u))^{2} \cdot  
 \frac{\sqrt{ u/\pi}}{\sqrt{2 + ut}}\prod_{m=1}^{\infty} \frac{1}{1 - e^{-2mt}\left( \frac{m-u}{m+u} \right)^{2}} \, . 
\end{equation}

We can fix the value of $N_{0}$ by comparing to the open string channel 
(see section \ref{open_chan}):
$N_{0} = 1/(\sqrt{2})$. It comes out that with this normalization the proportionality constant $C$ in 
(\ref{b_entr}) is equal to $1/\sqrt{2\pi}$. 
In addition to that we fix the normalization of Neumann boundary state for a coordinate $X$ compactified 
as $X\sim X + R$ by matching contributions of the zero modes as 
$$
\lim_{u\to \infty} \frac{1}{\sqrt{u}} \sim \frac{R}{\sqrt{2\pi }} \, . 
$$

With this normalization the full one-loop partition function (\ref{part_f}) for $D$ coordinates 
satisfying the boundary conditions 
(\ref{bc2}) and $26-D$ satisfying the Neumann ones reads 
\begin{eqnarray} \label{PF}
Z_{1}(u_{i}, a) &=& \frac{V_{26-D}}{(\sqrt{16\pi^{2}})^{26-D}} e^{-2a} \prod_{i=1}^{D} \Bigl[ 
(Z_{0}(u_{i}) )^{2} \cdot \nonumber \\ 
&& \int_{0}^{\infty}\frac{dt}{2\pi}
e^{2t} f^{D-24}(t)  \sqrt{ \frac{u_{i}}{ 8\pi  (1 + tu_{i}/2) } }  
\prod_{k=1}^{\infty} \frac{1}{1 - e^{-2mt}\left( \frac{m-u_{i}}{m+u_{i}} \right)^{2}} \Bigr] \, . 
\end{eqnarray}


\section{Green's function on the annulus, partition function and renormalization}
In the previous section we computed the one-loop partition function using computationally the shortest path. 
It is instructive to do an independent computation via Green's function on the annulus. In particular 
in this approach the renormalization implicitly hidden in the factor $(Z_{0}(u))^{2}$ becomes transparent. 

Consider an annulus $r_{0}\le |z| \le 1$. A Green's function $G(z, z')$ satisfying 
$$
-\frac{1}{4\pi} \frac{\partial}{\partial z} \frac{\partial}{\partial \bar z} G(z, z') = \delta^{2}(z, z')
$$
and boundary conditions (\ref{bc}) can be computed by exploiting an  ansatz corresponding to a decomposition into 
a particular solution and a general solution of the homogeneous equation represented in a form of Lorant expansion 
plus an additional ${\rm ln}|z|^{2}$ term: 
$$
G(z, z') = -{\rm ln} |z-z'|^{2} + \sum_{k=-\infty}^{+\infty} ( z^{k}f_{k}(z', \bar z') + \bar z^{k} 
\bar f_{k}(z', \bar z')) + C(z', \bar z'){\rm ln} |z|^{2} 
$$
where $f_{k}(z', \bar z')$ and  $ C(z', \bar z')$ are some unknown functions. 

Plugging in the ansatz into (\ref{bc}) and solving it we obtain 
\begin{eqnarray} \label{Green_fn}
&& G(z, z') = -{\rm ln} |z-z'|^{2} + \frac{2}{u} - 
\frac{2\left(1-\frac{u}{2}|z'|^{2}\right)\left(1-\frac{u}{2}|z|^{2}\right)  }{u(2-u{\rm ln} r_{0})} + \nonumber \\
&& \sum_{k\ne 0} [ (z\bar z')^{k} + (\bar zz')^{k}] \frac{k^{2} - u^{2}}
{k((k+u)^{2} - r_{0}^{2k}(k-u)^{2})} + \nonumber \\
&&  \sum_{k=1}^{+\infty} \Bigl[ \left( \frac{zr_{0}^{2}}{z'}\right)^{k} + 
\left( \frac{z'r_{0}^{2}}{z}\right)^{k} + \left( \frac{\bar zr_{0}^{2}}{\bar z'}\right)^{k} + 
\left( \frac{\bar z'r_{0}^{2}}{\bar z}\right)^{k}
\Bigr]\frac{(k-u)^{2}}{k((k+u)^{2} - r_{0}^{2k}(k-u)^{2})} \, . 
\end{eqnarray}
This Green's function can be explicitly checked to satisfy $G(z, z') = G(\frac{r_{0}^{2}}{\bar z}, z')$.

To determine the partition function $Z_{1}(u, r_{0})$ 
one can write out two kinds of equation corresponding to the variation of ${\rm ln} Z_{1}(u, r_{0})$ with 
respect to $u$ and the modulus $r_{0}$: 
\begin{equation} \label{eq1}
\frac {\partial {\rm ln} Z_{1}(u, r_{0})}{\partial u} = -\frac{1}{8\pi} \left( 
\int_{|z|=r_{0}}d\phi \langle X^{2}(z )\rangle + \int_{|z|=1}d\phi \langle X^{2}(z )\rangle \right) \, ,
\end{equation}  
\begin{equation} \label{eq2}
\frac {\partial {\rm ln} Z_{1}(u, r_{0})}{\partial r_{0}} = -\frac{r_{0}}{\pi (1- r_{0}^{2})} \int_{r_{0}\le |z| \le 1} 
d^{2}z\, \Bigl( \frac{T_{zz}}{\bar z^{2}} +  \frac{T_{\bar z\bar z}}{ z^{2}} \Bigr) 
\end{equation}
where $T_{zz}$ and $T_{\bar z \bar z}$ are the components of the stress-energy tensor. (See \cite{Callan2} 
for the derivation of the second equation. Note a factor of $-1/4\pi$ missed in their formula (4.11).)

The correlator $ \langle X^{2}(\phi )\rangle$ entering the first equation can be obtained by renormalizing 
the values of Green's function $G(z, z')$ on each of the boundaries in the limit when  $z$ approaches $z'$.
One finds that the divergences come from the bulk part $-{\rm ln} |z-z'|^{2}$ and from the terms in the second line 
of equation (\ref{Green_fn}). The last divergence is exactly of the same form as the first one and has a natural 
interpretation as a divergence coming from the image charge. Thus one can show that on each component of the boundary 
the divergent part of the two-point correlator is 
\begin{equation} \label{subtract}
lim_{\phi \to \phi'} G(\phi, \phi') = -2{\rm ln}| 1- e^{i(\phi - \phi')}|^{2} + \, \mbox{finite part} \, . 
\end{equation}
Subtracting the divergent part we obtain 
$$
\langle X^{2}(z )\rangle_{|z|=r_{0}} + \langle X^{2}(z )\rangle_{|z|=1} = 
\frac{4}{u}\Bigl[ \frac{1 - u{\rm ln}r_{0}}{2 - u{\rm ln} r_{0} }\Bigr] + 8\sum_{k=1}^{\infty} 
\frac{r_{0}^{2k}(k-u)(2k-u) - u(k+u) }{k((k+u)^{2} - r_{0}^{2k}(k-u)^{2}) } \, . 
$$
This expression can be now plugged into equation (\ref{eq1}) and integrated. This gives the partition function 
up to a factor that may depend on $r_{0}$. To fix this factor one may utilize the second equation (\ref{eq2}). 
Before explaining how this is done let us make some remarks on the subtraction we performed to make the
two-point correlator finite. 
First let us make an obvious remark   that  we employed the point splitting regularization. It is important 
to stay consistently with the same regularization when analyzing the system in various coordinates 
(see the sections below that deal with the open string channel). 
The corresponding counterterms are constants on  each boundary that are equal in value. Thus this subtraction is 
a renormalization of the coupling constant $a$. Of course this subtraction is the same as the one 
made in the tree level calculation \cite{Witten2}. In the calculation via the boundary state described in the 
previous section this regularization is implicitly present in the normalization factor $(Z_{0}(u))$ 
 of the boundary state  that enters squared (via two boundaries) in the partition function.
As we will see below the above subtraction is the only  subtraction 
logarithmic in scale that is needed to render the partition function finite. Therefore the beta functions 
(including the parts corresponding to the classical dimensions)  
for the coupling constants $a$ and $u$ are 
\begin{equation}
\beta_{u} = - u \, , \qquad \beta_{a} = -a -u \, .  
\end{equation}
In a more general case when one has boundary conditions (\ref{bc}) in $D$ directions one has 
\begin{equation}\label{beta}
\beta_{u_{i}} = - u_{i} \, , \enspace i=1, \dots D\, , \qquad \beta_{a} = -a -\sum_{i=1}^{D}u_{i} \, .  
\end{equation}

When using equation (\ref{eq2}) one first uses the Green's function (\ref{Green_fn}) to compute 
the components of the stress-energy tensor 
$$
T_{zz}(z) = lim_{z\to z'} \frac{1}{2} \Bigl[ \frac{\partial^{2}G(z, z')}{\partial z \partial z'} - 
\frac{1}{(z-z')^{2}} \Bigr] \, . 
$$
A straightforward computation yields 
$$
T_{zz} = \frac{1}{z^{2}}\Bigl[ -\frac{u}{4(2- u{\rm ln}r_{0})} - \sum_{k=1}^{\infty} 
\frac{r_{0}^{2k}k(k-u)^{2}}{(k+u)^{2} - r_{0}^{2k}(k-u)^{2}} \Bigr] \, . 
$$
Now we can plug this and the complex conjugated expression for $T_{\bar z \bar z}$ into (\ref{eq2}) 
and integrate it. After checking the result against the same kind of computation made using equation 
(\ref{eq1}) we obtain that up to an arbitrary overall numerical factor the partition function 
coincides with the one computed in the previous section (equation (\ref{part_f2})).
  

\section{One-loop bosonic  string divergences} 
In this section we remind the reader of the situation with one-loop divergences  
in the conformal case. Consider an open bosonic string satisfying 
Dirichlet boundary condition in $D$ directions and the Neumann ones the remaining  $26-D$ dimensions. 
Its one-loop partition function written in the open string sector has the form 
\begin{equation} \label{conf_PF}
Z_{1} = \int_{0}^{\infty} \frac{dT}{2T} {\rm Tr} e^{-2\pi T L_{0}} = 
 \frac{V_{26-D}}{ (\sqrt{ 8\pi^{2}\alpha' })^{26-D} } \int_{0}^{\infty} \frac{dT}{2T^{2}} T^{-(24-D)/2} e^{2\pi T} 
f(\pi T)^{-24} 
\end{equation}
This integral diverges at $T\to 0$. The divergent part in the trace comes from the open string tachyon states and 
is equal to 
\begin{equation}  \label{open_tach}
I = \frac{V_{26-D}}{(\sqrt{8\pi^{2} \alpha'})^{26-D}} \int_{R}^{\infty} \frac{dT}{2T} T^{-\alpha} e^{2\pi T}
\end{equation}
where we put in some cut-off $R$ and $\alpha = (26-D)/2$. The physical origin of this divergence is in the wrong 
sign of the tachyon mass squared. The divergent part can be represented as a point particle path-integral 
 (see Polchinski's book \cite{Polch_book} for a detailed discussion) 
$$
I \sim \int\frac{dl}{2l} \int dp^{26-D} e^{-(p^{2} - 1)l/2}  
$$
where $l$ is a proper time along the particle world line. 
This amplitude gives the (connected part of)  
one-loop vacuum amplitude for an open string  tachyon vibrating in $26-D$ dimensions. Equivalently in field theory  
the above formula can  be rewritten 
as a contribution of the tachyon to   the vacuum energy density (in space-time)  
$$
I \sim -\frac{1}{2}\int  \frac{dp^{26-D}}{(2\pi)^{26-D}}  {\rm ln}(p^{2} - 1)  
$$
that clearly indicates (by the presence of negative values under the logarithm) 
that the correctly defined amplitude may develop an imaginary part.

The divergent part (\ref{open_tach}) can be defined by means of analytic continuation. Here our 
discussion closely follows paper \cite{Marcus}.  One treats  
the exponential in (\ref{open_tach})  as a parameter $b$, evaluating the integral 
for $b = - 2\pi$ and then rotating in complex plane 
as $b\mapsto be^{-\pi i }$. The direction of rotation can be fixed by carefully inserting the $i\epsilon$ in the  
propagator (see \cite{Marcus}) for details). 
It is easy to obtain the imaginary part of the analytically continued integral using the formula 
\begin{equation} \label{an_cont}
{\rm Im} \int_{R}^{\infty}\frac{dx}{x}x^{-\alpha} e^{(b + i\epsilon)x} = \frac{\pi}{\Gamma(1 + \alpha)} b^{\alpha} \, .
\end{equation} 

This way we get for (\ref{open_tach}) 
\begin{equation} \label{Im_open_tach} 
{\rm Im} I = V_{26-D}  
\frac{\pi} {2 \Gamma \left( 14 - \frac{D}{2}\right) } 
\left( \frac{ |m^{2}_{\rm o. \, s. \,  tach.}|}{4\pi} \right)^{(26 - D)/2} \,  
\end{equation}
where $m^{2}_{\rm o. \, s. \,  tach.} = -1/\alpha'$ is the open string tachyon mass squared. 

The appearance of imaginary part for the partition function and thus for the one-loop space-time 
effective tachyon action of course signifies the instability of the system. It is known in field 
theory \cite{Coleman}, \cite{EWeinberg} that an imaginary part of the vacuum energy gives a decay rate 
of unstable vacuum. More precisely the decay rate per unit volume is $\Gamma = 2{\rm Im} E$.

One can also compute the real part of the analytically continued expression. However the real part 
will depend on the cutoff and needs to be pasted together with the other part of the partition function. 
The resulting number can be obtained numerically (see \cite{ Marcus} for an example of this type of computation).

The divergences due to closed string states can be studied by performing a 
closed/open string channel duality transformation 
$t=\pi/T$. The partition function (\ref{conf_PF}) takes the form  
\begin{equation}
Z_{1} = \frac{V_{26-D}}{ 2\pi (\sqrt{ 8\pi^{2}\alpha' })^{26-D} } \int_{0}^{\infty}dt t^{-D/2}  e^{2t} f^{-24}(t) \, . 
\end{equation}
When $D>2$ the divergent part comes only from the closed string tachyon states 
$$
I' = \frac{V_{26-D}}{ 2\pi (\sqrt{ 8\pi^{2}\alpha' })^{26-D} } \int_{R}^{\infty}dt t^{-D/2} e^{2t}   
$$
This expression corresponds to  a vacuum-vacuum tree-level diagram for the closed string tachyon  
$$
I' \sim \int \frac{dp^{26-D}}{(2\pi)^{26-D}} \frac{1}{p^{2} - 2} \, . 
$$

Using formula (\ref{an_cont}) we obtain for the imaginary part
\begin{equation} \label{Dirichlet2}
{\rm Im} I' = V_{26-D} \frac{ 2^{D/2 - 2}}{\Gamma(D/2)} 
\left( \frac{|m^{2}_{\rm c.\, s.\, tach.} |}{32\pi^{2}}\right)^{(26-D)/2} 
\end{equation}
where $m^{2}_{\rm c.\, s.\, tach.}= - 4/\alpha'$ is the closed string tachyon mass squared. 

When $D\le 2$ one also has a divergence corresponding to a dilaton tadpole. This is a true physical divergence 
that cannot be eliminated by analytic continuation and is due to the propagation of massless particles between 
the vacuum states. The momentum space propagator for a massless particle is $1/p^{2}$ that diverges on the mass
shell. The momentum conservation on the other hand requires that the massless particle emerges  from the 
vacuum with the zero momentum, i.e. precisely on the mass shell.  
The problem with this divergence  can be resolved by a shift of the dilaton background
 \cite{FS} that results in adding a source term to the space-time effective action.  


\section{Divergences in the closed string channel   in the presence of tachyon background} \label{CL_TAD}
In this section we consider divergences of the one-loop partition function (\ref{PF}) 
coming from the closed string sector.
By looking at the boundary state (\ref{b_state}) and the expression for the partition function (\ref{part_f})
we see that  divergences at $t\to 0$ come from the  vacuum part of the boundary state 
\begin{equation} \label{tachyon_st}
|{\rm tach.}\rangle \equiv N(u) \cdot \sqrt{u}  
exp\Bigl(- \frac{u\hat q^{2}}{4} \Bigr) |0\rangle 
\end{equation}
that describes a tachyon with a Gaussian wave function and in the case when $D\le 2$ from the part
\begin{equation} \label{dil_st}
|{\rm dil.} \rangle \equiv N(u) \cdot  \left( \frac{1-u}{1+u} \right) \cdot \sqrt{u}
  a_{1}^{\dagger} \tilde a_{1}^{\dagger}  exp\Bigl(- \frac{u\hat q^{2}}{4} \Bigr) |0\rangle 
\end{equation}
corresponding to a dilaton with the same Gaussian   wave function.

For $D\ge 3$ there is no divergence due to the dilaton (c.f. the discussion in the previous section). 
Physically one may think about this fact as follows. 
By plugging  the state  (\ref{dil_st}) into expression (\ref{part_f}) for the partition function 
we find that the contribution of this state up to an overall constant is 
$$ 
 I \sim \langle {\rm dil.} | \int_{0}^{\infty} dt \, e^{-\hat p^{2} t} \,  |{\rm dil.} \rangle = 
\langle {\rm dil.} | \frac{1}{\hat p^{2} } |{\rm dil.} \rangle = 
 \int d^{26}p \, \frac{1}{p^{2}} e^{-\sum_{i=1}^{D}\frac{p^{2}_{i}}{2u_{i}}} \, .   
$$
Fourier transforming this expression we obtain  
$$
I \sim \int d^{26}q \, \frac{1}{q^{24}}  e^{- \frac{1}{2} \sum_{i=1}^{D} u_{i}q^{2}_{i}  } 
$$ 
that can be interpreted as a Coulomb potential self-energy of a matter that has a Gaussian density 
distribution  in $D$-directions. For $D>3$ it is finite. The Gaussian factor plays a role of 
effective infrared regulator. 
However in the limit $u\to 0$ the divergence 
will show up as a more singular behavior of the partition function.

 Let us postpone a further  investigation of  the dilaton divergences  until section \ref{potential} 
 and consider now
divergences due to the tachyon state. The divergent contribution of state (\ref{tachyon_st}) to the 
partition function has the form 
$$
 I_{\rm cl. \, tach.} (u)= \frac{V_{26-D}}{(\sqrt{16\pi^{2}})^{26-D}} e^{-2a}\prod_{i=1}^{D}\Bigl[ 
 (Z_{0}(u_{i}) )^{2}   \int_{0}^{\infty}\frac{dt}{2\pi}
e^{2t}   \sqrt{ \frac{u_{i}}{ 8\pi  (1 + tu_{i}/2) } } \Bigr] \, . 
$$
For the sake of simplifying the computation we will assume that all $u_{i}$'s are equal to a single parameter $u$. 
Then changing the exponent $2$ to a complex parameter $-b$ and  shifting 
 the integration  variable we can rewrite the above integral  as 
$$
I_{\rm cl. \, tach.}(u,  b)  = \frac{V_{26-D}}{(\sqrt{16\pi^{2}})^{26-D}} (e^{-a}(Z_{0}(u))^{D})^{2} 
\frac{1}{2\pi (4\pi)^{D/2}} e^{2b/u}\int_{2/u}^{\infty} dt'\, \frac{e^{-bt'}}{(t')^{D/2}}  \, . 
$$
Applying formula (\ref{an_cont}) we obtain 
\begin{equation} \label{im_cl}
{\rm Im} I_{\rm cl. \, tach.} (u) =  \frac{V_{26-D}}{(\sqrt{16\pi^{2}})^{26-D}} (e^{-a}(Z_{0}(u))^{D} )^{2} 
\frac{1}{ (4\pi)^{D/2}} e^{-4/u} \frac{ 2^{D/2-2}}{\Gamma(D/2)} \, . 
\end{equation}
As we will show in section \ref{tension} minimizing the space-time effective action with respect to $a$ 
the limit of the quantity  $ e^{-a}(Z_{0}(u))^{D} $ along the RG trajectory  $u\to \infty$, $a\to \infty$ 
is a number $(\sqrt{2\pi})^{D}$ just as in the tree level computation \cite{Kutasov1}. 
Thus the imaginary part (\ref{im_cl}) as $u=\to \infty$ approaches the finite value 
\begin{equation} \label{uinf}
{\rm Im} I_{\rm cl. \, tach.} (\infty) = 
\frac{V_{26-D}   (2\pi )^{D/2}    } { 4 (\sqrt{16\pi^{2}})^{26-D}  \Gamma(D/2)  } \, .  
\end{equation} 

 Up to  a numerical factor 
(that essentially comes from $(e^{-a}Z_{0}(u) )^{2D} $ ) ((\ref{uinf}) agrees with (\ref{Dirichlet2}) 
and we see that for this divergence  the naive prescription of first taking the $u \to \infty$ limit that 
yields the Dirichlet boundary conditions and then performing the analytic continuation gives the same result 
as the accurate procedure described above.  We see however that along the RG flow the imaginary part 
due to 
the closed string tachyon grows (in absolute value) from zero at $u=0$ to the constant value (\ref{uinf}).

The closed string tachyon one-loop divergence in the presence of the quadratic open string tachyon background 
was also considered in \cite{Nakamura} where  a different treatment than analytic continuation is proposed.  

\section{Open string channel} \label{open_chan}
The divergences of the partition function (\ref{PF}) in the limit $t\to 0$ are best studied 
in the open string channel that describes an open  string propagating in a  (Euclidean) time loop of length 
$2\pi T = 2\pi^{2}/t$ and satisfying the  boundary conditions (\ref{bc3}) that for the sake of readers 
convenience we write here once more:
\begin{equation} \label{bc3'}
-\frac{\partial X}{\partial \sigma} + \frac{u}{T}X = 0\, , \enspace \sigma=0 \, , \qquad 
\frac{\partial X}{\partial \sigma} + \frac{u}{T}X = 0\, , \enspace \sigma = \pi \, . 
\end{equation}
(To simplify the notations in this section we will drop the primes at the open string world sheet 
coordinates $\sigma', \tau'$.)  
 Although these boundary conditions have a nonlocal time-dependence via the factors of $\frac{1}{T}$ 
in order to compute the partition function 
\begin{equation} \label{open_pf}
Z_{1} = \int_{0}^{\infty} \frac{dT}{2T} {\rm Tr} e^{-2\pi T H(u/T)} 
\end{equation}
we may first canonically quantize the theory with the boundary conditions 
\begin{equation} \label{bc3''}
-\frac{\partial X}{\partial \sigma} + vX = 0\, , \enspace \sigma=0 \, , \qquad 
\frac{\partial X}{\partial \sigma} + vX = 0\, , \enspace \sigma = \pi  
\end{equation}
where $v$ is a constant,
write down the partition function and then replace $v$ with $u/T$. 
The Hamiltonian $H(v)$ will have a spectrum of the form 
$$
h = E_{Cas}(v) + \sum_{n=0}^{\infty} N_{n} \lambda_{n}(v) 
$$
where $E_{Cas}(v)$ is the Casimir energy in the background (\ref{bc3''}),   
$\lambda_{n}(v)$ are oscillator frequencies and $N_{n}$ - the occupation numbers. 
The partition function (\ref{open_pf}) is thus of the form 
\begin{equation}\label{open_pf2}
Z_{1} = \int_{0}^{\infty} \frac{dT}{2T} e^{-2\pi T E_{Cas}(u/T)}
\prod_{n=0}^{\infty} \frac{1}{1 - e^{-2\pi T \lambda_{n}(u/T)}} 
\end{equation}

If the spectrum in the presence of $u$- background is not significantly modified the divergences 
in (\ref{open_pf2}) will come only from a finite number of the first excited states, i.e. 
the ground state with energy $E_{Cas}(u/T)$ and possibly some number of the low lying  excited states.

We start by deriving the spectrum of frequencies. A general solution  to the Laplace equation 
$$
\Bigl( \frac{\partial^{2}}{\partial \tau^{2}} +  \frac{\partial^{2}}{\partial \sigma^{2}} \Bigr) 
X = 0 
$$ 
with the boundary conditions (\ref{bc3''}) can be expanded into eigenfunctions as 
$$
X(\sigma, \tau) = \sum_{n=-\infty}^{+\infty} e^{\lambda_{n}\tau} ( \alpha_{n} e^{-i\sigma \lambda_{n}} + 
\beta_{n} e^{i\sigma \lambda_{n}} )\, . 
$$  
By plugging in this expansion into the boundary conditions (\ref{bc3''}) we obtain the system of two 
linear homogeneous equations. The corresponding determinant factorizes as 
$$
\Delta = \left( {\rm tan} \left( \frac{\pi\lambda_{n}}{2} \right) - \frac{v}{\lambda_{n}} \right) 
\left( {\rm cot} \left( \frac{\pi\lambda_{n}}{2} \right) +  \frac{v}{\lambda_{n}}   \right) \, .  
$$  

Thus we obtain two spectral equations 
\begin{eqnarray} \label{spectrum}
&&  {\rm tan} \left( \frac{\pi\lambda_{n}}{2} \right) =  \frac{v}{\lambda_{n}} \, , \nonumber \\
&& {\rm cot} \left( \frac{\pi\lambda_{n}}{2} \right) = -   \frac{v}{\lambda_{n}} 
\end{eqnarray}
defining the even and odd parity eigenvalues respectively. We would like to remark here that the 
spectral problem we are looking at here is equivalent to the following quantum mechanical model. 
Consider a one-dimensional system on a circle with coordinate  $0\le x < 2\pi$ subject to 
${\bf Z}_{2}$ orbifold identification $x\sim 2\pi - x$. If we consider a quantum mechanical 
particle in this space with a potential   
$$
V(x) = 2u \delta(x) + 2u\delta(x-\pi) 
$$ 
then  for the corresponding Schroedinger operator one obtains exactly the same spectrum as (\ref{spectrum}) 
provided one restricts himself to the  solutions that are even under the reflection $x\to x-2\pi$.

Alternatively both parity branches can be combined in a single equation 
$$
{\rm tan}(\pi \lambda_{n}) = \frac{2\lambda_{n}v}{\lambda^{2}_{n} - v^{2}} \, . 
$$
It is clear from this equation or equations (\ref{spectrum}) that  for sufficiently large $n$ one 
recovers the regular Regge spectrum with the first correction of the form 
\begin{equation} \label{approx_spec}
\lambda_{n} = n + \frac{2v}{\pi n} + {\cal O}( (v/n)^{2}) \, .  
\end{equation} 

For $v<1$  equation (\ref{approx_spec}) is a fairly good approximation to all eigenvalues with the 
exception of the lowest eigenvalue. The last one is approximately 
\begin{equation} \label{lambda0}
\lambda_{0} \approx \sqrt{\frac{2v}{\pi} }  
\end{equation}
with the correction being of the order of $v^{3/2}$. 

Despite the fact that we cannot solve the transcendental equations (\ref{spectrum})  exactly (and thus 
cannot use the direct mode summation formula) we still will be able to derive an integral formula 
for $E_{Cas}(v)$ valid for all (nonnegative) values of $v$. Before we go into that let us first consider as  
a warmup the small $v$ case when  the expressions (\ref{approx_spec}), (\ref{lambda0}) provide a good 
approximation for the spectrum. 
Using those expressions and the mode summation method we can compute an approximate expression for the 
 Casimir energy that is valid  up to the terms depending on $v$ as $v^{p}$, $p>1$. Note that after we substitute  
$v= u/T$ and plug these terms into the Hamiltonian we obtain  factors of the form 
$e^{-2\pi u^{p}T^{1-p}}$ that tend  to 1 as $T\to \infty$. 
Thus for small $u$  our approximation 
 will allow us  to derive the leading divergence of the partition function as $T\to \infty$.   
Moreover, despite its seemingly limited range of use, this asymptotics essentially  sets the 
imaginary part of the analytically continued $T\to \infty$ divergence. This is due to the following observation. 
The imaginary part should not depend on the lower cutoff in $T$-integration. Thus for any value of 
$u$ we can place the cutoff high enough so that $v=u/T$ is small and thus the approximation we are talking about is 
useful.

To compute the Casimir energy in such a way that the resulting partition function will match 
the one computed in the closed string channel  we must employ the equivalent regularization scheme. 
Thus we should proceed by using  the point splitting regularization. Note that there is a 
subtlety here. The point splitting parameter $\phi' - \phi = \epsilon$ used on the annulus 
(\ref{subtract}) corresponds to a point splitting $\epsilon' = \epsilon\cdot T$ on the strip in the 
open channel. The factor $T$ comes from the coordinate mapping 
$$
z = e^{-\frac{ \sigma + i \tau}{T}}
$$  
relating the two pictures. Here $z$ is the complex coordinate on the annulus and $\sigma, \tau$
are coordinates on the strip.

With this identification in mind we can write the point slitting regulated Casimir energy for small 
values of $v$ as 
\begin{eqnarray*}
&& E_{Cas}(v) = \frac{1}{2}\lambda_{0} + \frac{1}{2}\sum_{n=1}^{\infty} \lambda_{n}e^{-\epsilon'\lambda_{n}} = 
\sqrt{\frac{v}{2\pi} } +   \frac{1}{2}\sum_{n=1}^{\infty}
\left( n +  \frac{2v}{\pi n}\right) e^{-\epsilon'(n + 2v/(\pi n))} + o(v) = \\
 && \sqrt{\frac{v}{2\pi} } -\frac{1}{24} + \frac{1}{2T^{2}\epsilon^{2}} - \frac{v}{\pi} 
-\frac{v}{\pi}{\rm ln}\epsilon - \frac{v}{\pi}{\rm ln}T  + o(\epsilon) 
+ o(v) 
\end{eqnarray*}
Subtracting the divergent quadratic  and  logarithmic parts  
 we obtain after sending $\epsilon \to 0$
\begin{equation} \label{Cas_en}
E_{Cas}(v) = -\frac{1}{24} + \sqrt{\frac{v}{2\pi} } - \frac{v}{\pi} - \frac{v}{\pi}{\rm ln} T + o(v)
\end{equation}

Therefore it follows that  
\begin{equation} \label{asymptotics}
lim_{T\to \infty} {\rm Tr} e^{-2\pi T H(u/T)} = T^{2u}e^{2u} 
exp\left( \frac{\pi T}{12} - 2\sqrt{2\pi u T} -2\pi T\cdot f(u/T) \right)  
\end{equation}
where $f(u)$ is some unknown function that has the property  
$$
lim_{T\to \infty} T\cdot f(u/T) = 0 \, .
$$
We independently obtained the same asymptotics (up to an overall exponent $e^{Cu}$ for which we were unable to pin down 
the value of $C$) by analyzing the expression for the partition function in the closed string sector  
by means of the  Euler-Maclaurin summation formula. 


We would like to derive now a general integral formula for the Casimir energy $E_{Cas}(v)$ that is valid for 
all values of $v$. This can be done as follows. 
 It is easy to show that the spectrum $\lambda_{n}$ 
   coincides (except for the zero poit) with  the set of 
zeroes of an entire analytic function  
$$
\phi(z) = e^{i\pi z}(v + iz)^{2} - e^{-i\pi z}(v - iz)^{2} \, .
$$

The regulated Casimir energy then can be represented as a contour integral 
\begin{equation}
E_{Cas}(v, \epsilon')= \frac{1}{4\pi i} \oint z e^{-\epsilon' z} d\, {\rm ln}\phi(z) 
\end{equation}
where the contour  should encircle  the positive eigenvalues and can be conveniently chosen 
to consist of two slanted rays: $z = (i + \delta) x$, $z = (-i + \delta) x$ where $\delta>0$ 
and $x$ runs from zero to infinity. One should also keep in mind that we do not include the zero 
into the spectrum. Although formally it does not contribute to  the regulated infinite sum of the eigenvalues 
one may still wish to avoid potential troubles by modifying the contour by cutting out small initial segmets 
of the rays and connecting their endpoints by a small half-arc. 

To  simplify the manipulations below we note  that both  $\delta$ and the point splitting 
parameter $\epsilon'$ act as regulators, and one can achieve the same result by keeping only $\epsilon'$ and assuming that 
it has an appropriate imaginary part that provides a damping exponential factor.  
 With this in mind and taking also into account that the function $\phi(z)$ is odd we can 
write $E_{Cas}(v, \epsilon')$ as the following integral 
$$
E_{Cas}(v, \epsilon') = - \frac{1}{2\pi} \int_{0}^{\infty} x cos(\epsilon' x)
d\, {\rm ln}\Bigl( e^{\pi x} (v + x)^{2} - e^{-\pi x}(v- x)^{2} \Bigr) \, . 
$$
Factoring out the term $ e^{\pi x} (u + x)^{2}$ we obtain after one partial integration 
\begin{eqnarray*}
&& E_{Cas}(v, \epsilon') =  \frac{1}{2\pi} \int_{0}^{\infty} 
{\rm ln}\Bigl( 1 - e^{-2\pi x}\left( \frac{ x-v}{x+v } \right)^{2}  \Bigr) 
d(xcos(\epsilon' x)) \\ 
&& -\frac{1}{2}\int_{0}^{\infty} x cos(\epsilon' x) dx - 
\frac{1}{\pi} \int_{0}^{\infty} \frac{x cos(\epsilon' x) }{x + v} dx \, . 
\end{eqnarray*} 
Here the first integral is finite as $\epsilon' \to 0$, the second term is quadratically 
divergent and the third one is  logarithmically divergent. 
The divergences are regulated by imaginary part of $\epsilon'$ 
that is chosen with an appropriate sign. 
Let us show how to treat  the logarithmic divergence.  We can rewrite the logarithmically divergent part $I_{log}$ as 
$$
I_{log} = \frac{v}{\pi} e^{v\epsilon'}(-E_{\rm i}(-v\epsilon'))
$$  
where $E_{\rm i}(x)$ denotes the exponential integral function. 
Its asymptotics near zero is such that (see for example \cite{Magnus}) $-E_{\rm i}(-x) = -\gamma - {\rm ln}(x) + o(x)$, 
$|arg x|<\pi$ where $\gamma$ is the Euler constant. 
Using this asumptotics we can go ahead subtract  the infinities 
end send $\epsilon = \epsilon'/T$ to zero.  We obtain the following expression 
\begin{equation} \label{ECAS}
 E_{Cas}(v) = \frac{1}{2\pi} \int_{0}^{\infty} 
{\rm ln}\Bigl( 1 - e^{-2\pi x}\left( \frac{ x-v}{x+v } \right)^{2}  \Bigr) dx 
- \frac{v}{\pi}{\rm ln}(vT) - \frac{v}{\pi} \gamma \, . 
\end{equation}
Note that the first term in this expression we 
could have easily gotten starting from the closed string expression (\ref{PF}) and 
using the Euler-Maclaurin summation formula. Thus we see that in terms of that formula 
the open string channel Casimir energy is essentially the integral approximation to the infinite 
series while  the corrections (that can be written for example as integrals involving 
saw-tooth function and its integrals) correspond to the excited states. 
Note that the first term in (\ref{ECAS}) gives the standard conformal Casimir 
energy $-\frac{1}{24}$ in the limits $v\to 0$ and $v\to \infty$.


Let us discuss now  the divergence $T\to \infty$ ($t\to 0$) of the partition 
function. The general form of asymptotics is given by formula (\ref{asymptotics}) 
in which the function $f(x)$ is in principle extractable from our general expression 
(\ref{ECAS}). This function cannot be dropped because the corresponding contributions 
are still divergent. 
Also  note that 
the states coming from the excited levels of the first oscillator with frequency $\lambda_{0}$ also 
have negative energy and thus contribute to the divergence (\ref{asymptotics})  an overall  factor 
$$
\frac{1}{1- e^{-2\pi T\lambda_{0}(u/T)}} \approx \frac{1}{1- e^{-2\sqrt{2\pi Tu }}} \, . 
$$
Note that in the limit $u\to 0$ this factor restores the contribution of zero modes 
$$
\frac{1}{2\sqrt{2\pi Tu }} \, . 
$$
By comparing the $u\to 0$ behavior of the partition function in the closed string channel (\ref{part_f2}) 
to the expression  above we find  the normalization of the boundary state (\ref{b_state}).

Thus we can write down the following general expression for 
the tachyon divergent part of the partition function for $26$ scalars with $D$ scalars satisfying the boundary 
condition (\ref{bc3'}) has the form
\begin{eqnarray} \label{open_tp}
 I_{\rm o. \, tach.}(u) &=& e^{2u} \frac{V_{26-D}}{(16\pi^{2})^{(26-D)/2}}  
 \int_{R}^{\infty} \frac{dT}{2T}  
 T^{2u} T^{-(26-D)/2}  \cdot \nonumber \\
&& exp\left(2\pi T + 2D\sqrt{2\pi u T} -2\pi DT\cdot f(u/T) \right) 
  \frac{1}{1- e^{-2D\sqrt{2\pi Tu} -2D\pi T \phi(u/T)}}
\end{eqnarray}
where $\phi(u)$ stands for a correction to $\lambda_{0}(u)$, $b=2\pi$ to be rotated to $-2\pi$, and $R$ is 
a cutoff.  Further analysis of this formula is obstructed by two facts: our integrable representation 
(\ref{ECAS}) is not very useful in analytic manipulations and second, we do not have any explicit 
analytic  expression  for $\phi(u)$. 

The terms contained in $f(u)$ and $\phi(u)$ are responsible for the  flow 
from the Neumann to the Dirichlet spectrum. In the course of that flow  the even integer eigenvalues get 
shifted by one unit and become  the odd ones and vice versa.
Since the limit $u\to \infty$ sends the variable $u/T$ to infinity and the limit $T\to \infty$ sends it to zero 
for a fixed cutoff $R$ the contributions of $f$ and $\phi$ become more and more important in that limit.
However as we argued above for the purposes of computing the imaginary part of the analytic continuation 
one may always adjust the lower cutoff so that the terms in $f$ in $\phi$ are all subdominant. 
One can expand in this terms so that the typical expression $u$-dependent term in the integral  is of the form 
\begin{equation} \label{O}
O(u) \cdot \int_{R}^{\infty} e^{-bT + 2D\sqrt{2\pi u T}}T^{\alpha(u)}
\end{equation} 
where $\alpha(u)$ and $O(u)$ are such that   $\alpha(u)\to constant$, $O(u) \to 0$ as $u\to 0$. 
Each term of this form can be analytically 
continued in $b$. 

One may be interested in two kinds of questions about the analytic continuation and the imaginary part. 
First one may worry whether 
taking the limits $u\to 0$ $u\to \infty$ commutes with analytic continuation, or in other words do we recover 
the standard analytically continued Dirichlet and Neumann partition functions. If this were not true it 
would clearly signal some inconsistency of the analytic continuation procedure applied to the off shell 
situation at hand. In view of expansion (\ref{O}) above it seems to us that this is not the case.

A second, less formal question one may be interested in has to do with the physical interpretation of the imaginary part.  
Since it  gives the decay rate of the unstable vacuum,
it could be used as a measure of  the vacuum stability. It would be very interesting then to see how it behaves  along 
the RG trajectory.  One  may expect that the flow monotonically decreases to the value set by    (\ref{Im_open_tach}) 
for the appropriate  $D$. Unfortunately we did not gain enough analytical control over the open string channel 
   to see that.  


\section{Correction to the tachyon potential} \label{potential}
Consider now a boundary perturbation specified by coupling constants $u_{i}$ switched on in $D$ directions with 
the other $26-D$ Neumann directions being compactified as $X_{i}\sim X_{i} + R_{i}$, $i=D+1, \dots, 26$. 
We will restrict ourselves to the case $D\ge 3$ so we will not have to deal with the dilaton divergence directly 
from the start, although inevitably it will show up in the $u\to 0 $ limit. 

We assume that the space-time one-loop effective action is given by the expression 
$$
S = (1 + \beta_{a}\frac{\partial}{\partial a} + \sum_{i=1}^{D}\beta_{u_{i}}\frac{\partial}{\partial u_{i}} ) 
(\frac{1}{g}Z_{0}(u, a) + Z_{1}^{a.c.}(u, a) )  
$$ 
where $Z_{0}(u, a) = e^{-a}\prod_{i=1}^{D} Z_{0}(u_{i})\prod_{i=D+1}^{26} \frac{R_{i}}{\sqrt{4\pi }}$ 
is the tree-level partition function, $g$ is a string coupling 
constant  and 
$$
 Z_{1}^{a.c.}(u, a) = e^{-2a} Z_{1}^{a.c.}(u_{1}, \dots , u_{D}) 
$$
stands for the one-loop partition function (\ref{PF}) defined by means of analytic continuation. For $D\ge 3$ it takes finite 
values and contains an imaginary part due to both closed and open string tachyons. 

At this point we would like to restore the factors of $\alpha'$. This can be easily done by the substitutions:
$u\mapsto u\alpha'/2$, $R_{i}\mapsto R_{i}\sqrt{2/\alpha'}$. 
Substituting the beta functions (\ref{beta}) we obtain 
\begin{eqnarray} \label{S_ws}
S &=& \frac{1}{g}e^{-a}(a + \sum_{i} \frac{\alpha'}{2}u_{i} - u_{i}\frac{\partial}{\partial u_{i}} + 1) 
\prod_{i} Z_{0}(u_{i})\prod_{i=D+1}^{26} \frac{R_{i}}{\sqrt{2\pi \alpha' }} + 
\nonumber \\ 
&& e^{-2a} (2a + \sum_{i} \alpha'u_{i} - u_{i}\frac{\partial}{\partial u_{i} } + 1)Z_{1}^{a.c.}(u_{1}, \dots , u_{D}) \, . 
\end{eqnarray}

On the other hand this expression  should coincide with the space-time action 
\begin{equation} \label{space-time}
S = T_{25} \int d^{26} X \, [ f(T) \partial_{i} T \partial^{i} T + V(T) + {\rm higher \enspace derivative \enspace terms} ]
\end{equation} 
evaluated on the quadratic tachyon profile
\begin{equation}\label{profile}
T(X) = a + \frac{1}{4}\sum_{i} u_{i} X_{i}^{2} \, . 
\end{equation}
The tree level computation \cite{Gerasimov1}, \cite{Kutasov1} gives 
$$
f(T) = e^{-T} \, , \qquad V(T) = e^{-T}(1 + T)      
$$
and the value of $T^{tree}_{25} = \frac{1}{g}(2\pi \alpha')^{-13}$ where we inserted the $1/g$ factor to match with our 
conventions. Our considerations below follow more closely paper \cite{Kutasov1}. 
Note that our normalizations are slightly different from those in \cite{Kutasov1}. Their coupling constant $u$ is 
$4$ times our $u$.

Let us now first study the expression (\ref{S_ws}) in the limit $u\to 0$. 
We can  represent $Z_{1}^{a.c.}$ as a sum of the ``bulk part'' that comes from integration over a region of moduli 
space with cutoffs on both ends, and the analytically continued ``tails'' coming from the two boundaries of the 
moduli space. As $u$ goes to zero the bulk part tends to the bulk part of the Neumann partition function. 
The  contributions  to the tails due to open and closed string tachyons were discussed in the previous sections. 
We argued that both analytically continued tails due to tachyons  tend to the corresponding tachyon contributions 
of the string with Neumann boundary conditions. The only problem in the $u\to 0$ comes from the dilaton states. 

For simplicity let us assume $u_{1}=\dots = u_{D}=u$.
Then the dilaton contribution to the partition function that hides in it the usual dilaton tadpole has the form 
\begin{eqnarray}
I_{\rm dilaton}(u, a) &=& (e^{-a}(Z_{0}(u))^{D})^{2} \frac{V_{26-D}}{(\sqrt{16\pi^{2}})^{26-D}} \cdot 
( (24- D) + D\left(\frac{1-u}{1+u}\right)^{2}) \cdot \nonumber \\ 
&&  \int_{0}^{\infty}\frac{dt}{2\pi} \frac{1}{((2/u + t)4\pi)^{D/2}} \, . 
\end{eqnarray}  
Evaluating the integral we obtain an asymptotics of the form 
$$
I_{\rm dilaton}(u) \sim Const\cdot \frac{1}{u} \cdot \left( \frac{1}{\sqrt{u}} \right)^{D} \, , \enspace u\to 0 
$$
that contains an extra $1/u$ factor standing at the usual volume element. 

In the space-time effective action this term  can be  represented by a nonlocal term of the form 
$$
\frac{1}{\partial^{2}}e^{-2T}(1 + 2T)  \, . 
$$  
Thus as $u\to 0$ we have an explicit infrared problem in our action. It must be resolved by introducing a new massless 
degree of freedom about which we of course were aware from the very beginning - the  dilaton field. 

In this paper we will not try to  incorporate the dilaton field in our action. Instead we will stay at small 
but finite $u$ (we may think that we put our system in a box) and find a  correction to the tachyon 
potential.

Let us come back now to  the consideration of   background $u_{i}$ (where $u_{i}$'s are not necessarily equal to each other) 
  for $D$ directions 
where we assume $D\ge3$ and all $u_{i}$'s are away from zero. 
As before  the remaining $26-D$ directions  are compactified in a box with sides $R_{i}$.
 To derive the correction to the  potential we have to match  the terms 
at $1/\sqrt{u}$ as $u\to 0$ in the expression coming from the partition function (\ref{S_ws}) 
and the one coming from the space-time action (\ref{space-time}). The partition function (\ref{PF}) being plugged 
into the second term in (\ref{S_ws}) gives 
\begin{equation} \label{1}
{\cal S}^{(1)} \sim \frac{V_{26-D}}{(\sqrt{\alpha' \pi})^{26-D}} e^{-2a}(1 + \frac{D}{2} + 2a) \frac{1}{(\sqrt{8\pi})^{26}}
\Bigl(\prod_{i=1}^{D}\frac{1}{\sqrt{u_{i}\alpha'/2}}\Bigr) \int_{0}^{\infty} \frac{dt}{2\pi}e^{2t}f^{-24}(t)
\end{equation}
where $V_{26-D} = \prod_{i} R_{i}$ and we restored the factors of $\alpha'$. 
On the other hand by inserting the quadratic profile (\ref{profile}) into the potential density in (\ref{space-time}) 
that has to be of the form 
$$
k\cdot e^{-2T}(1 + 2T) 
$$
where $k$ is some constant, we obtain 
\begin{equation} \label{2}
{\cal S}^{(1)} \sim T_{25}k V_{26-D} e^{-2a}(1 + \frac{D}{2} + 2a) \prod_{i=1}^{D} \sqrt{  \frac{2\pi}{u_{i}}} \, . 
\end{equation} 
Equating (\ref{1}) and (\ref{2}) we obtain 
 the following expression for the 1-loop effective potential.  

\begin{equation} \label{poten}
V_{1 \rm loop}(T) =  e^{-T} (1 + T)  + g Z e^{-2T} (1 + 2T)  
\end{equation}
where $Z$ is up to normalization the analytically continued Neumann partition function with the dilaton tadpole subtracted 
which we can formally write as 
$$
Z = \frac{1}{(\sqrt{4\pi})^{26}}  \int_{0}^{\infty} \frac{dt}{2\pi} e^{2t}f^{-24}(t) \, . 
$$ 
Note that the procedure of matching (\ref{1}) and (\ref{2}) is quite sensitive to the normalization of the partition function
(\ref{PF}). In particular if we had a different normalization of the zero modes for the Neumann boundary conditions we would 
find that the coefficient $k$ in (\ref{2}) depends on $D$ which is  clearly   an inconsistency. The fact that we obtained the 
correct normalization based on purely world sheet considerations seems to be quite encouraging.

From the above expression for the correction to the potential we can read off the one-loop corrected D25 brane tension
\begin{equation} \label{T25}
T_{25}^{1l} = T_{25}^{tree} ( 1 + g Z) \, . 
\end{equation}

The imaginary part of Z is due to the open string tachyon and is given by the expression  (\ref{Im_open_tach} ) for $D=0$, 
multiplied by $1/(4\pi)^{13}$. In the proper definition of the tension (\ref{T25}) one should consider only the 
real part. The imaginary part specifying the decay rate is a separate piece of information. See \cite{EWeinberg} 
for a detailed discussion of how it works in field theory.


\section{Loop corrected Dp-brane tensions} \label{tension}
In this section we would like to study the ratio of the brane tensions by finding the limiting value of 
the effective action (\ref{S_ws}) in a similar way to how it was done in \cite{Kutasov1}. 

We begin as in \cite{Kutasov1} by extremizing the action $S(a, u_{i})$ with respect to $a$. 
To simplify the formulas  let us assume that $Z_{0}(u)$  denotes the complete tree level partition function 
for $D$ boundary conditions with $u_{i}$ and the $D-26$ Neumann ones.
Differentiating the expression (\ref{S_ws}) with respect to  $a$ and equating the result to zero we obtain 
the following equation for $a^{*} = a^{*}(u)$ 
\begin{eqnarray*} 
Z_{0}(u) + 2ge^{-a^{*}}Z_{1}(u) &=& (a^{*} +  \sum_{i} \frac{\alpha'}{2}u_{i} - u_{i}\frac{\partial}{\partial u_{i} } + 1) 
Z_{0}(u) + \nonumber \\ && 
2ge^{-a^{*}} (2a^{*} + \sum_{i} \alpha'u_{i} - u_{i}\frac{\partial}{\partial u_{i} } + 1)Z_{1}^{a.c.}(u) 
\end{eqnarray*}
Let us represent now $a^{*}$ as a sum $a^{*} = a_{0}^{*} + \Delta a^{*}$ where 
$$
a^{*}_{0} = -  \sum_{i} \frac{\alpha'}{2}u_{i} + \sum_{i} u_{i}\frac{\partial}{\partial u_{i}} {\rm ln} Z_{0}(u) 
$$
is the tree level solution. 
Let us also use the fact that $Z_{1}^{a.c.}(u)$ has the form $Z_{1}^{a.c.}(u) = Z^{2}_{0}(u)\tilde Z_{1}(u)$ (\ref{PF}). 

Then we have the following equation for the correction term $\Delta a^{*}$
$$
-\Delta a^{*} = g \left( e^{-a_{0}^{*}}Z_{0}(u) \right) e^{- \Delta a^{*}}
(4\Delta a^{*} -2 \sum_{i} u_{i}\frac{\partial}{\partial u_{i}} )\tilde Z_{1}(u)\, .   
$$
Note that as $u\to \infty$ the factor $\left( e^{-a_{0}^{*}}Z_{0}(u) \right)$ monotonically 
decreases to the value $(\sqrt{2\pi})^{D}$. The above equation for $ \Delta a^{*}$ is exact. 
However it is more consistent, in view of higher loop corrections, to keep in it only terms of 
the first order in $g$. This gives 
$$
\Delta a^{*} = 2g\left( e^{-a_{0}^{*}}Z_{0}(u) \right)
\sum_{i} u_{i}\frac{\partial}{\partial u_{i}} )\tilde Z_{1}(u) \, . 
$$
It follows from  this equation that 
$lim_{u\to \infty}\Delta a^{*} = 0$ if the following assumptions on $\tilde Z_{1}(u)$ are true: 
$$
lim_{u\to \infty} \tilde Z_{1}(u) = Const \, , \qquad lim_{u\to \infty} \sum_{i} u_{i}\frac{\partial}{\partial u_{i}} 
\tilde Z_{1}(u) = 0 \, . 
$$
Both assumptions can be easily shown to be true for the ``bulk'' part of the  $\tilde Z_{1}(u)$, i.e. for 
the part where integration over the modulus has cutoffs on the two ends. Moreover it follows from our considerations
 in section \ref{CL_TAD} that these  assumptions are also true for the contribution of the closed string channel boundary 
($t\to \infty$, $T\to 0$). As for the open string channel part we have to leave it at the level of conjecture. 

With these assumptions being true we can safely plug in  the tree level solution $a^{*} = a^{*}_{0}$ into 
equation (\ref{S_ws}) and take the limit $u\to \infty$. 
We obtain (restoring the explicit volume factor at the tree level partition function)  
\begin{eqnarray*} 
&& \lim_{u\to \infty} S(u, a^{*}(u)) = 
\lim_{u\to \infty}  \Bigl[ e^{-a^{*}(u)}Z_{0}(u)  \prod_{i=D+1}^{26}  \frac{R_{i}}{\sqrt{2\pi \alpha'}}
+ \\ 
&& g(e^{-a^{*}(u)}Z_{0}(u))^{2} \tilde Z_{1}(u)  \prod_{i=D+1}^{26}  \frac{R_{i}}{\sqrt{\pi \alpha'}} \Bigr]   = \nonumber \\
&&  (2\pi)^{D/2} \prod_{i=D+1}^{26}  \frac{R_{i}}{\sqrt{2\pi \alpha'}}
  + g(2\pi)^{D} \frac{1}{(\sqrt{8\pi})^{26}} (\sqrt{2})^{D}  \int_{0}^{\infty}\frac{dt}{2\pi} t^{-D/2} e^{2t}f^{-24}(t) 
\prod_{i=D+1}^{26}  \frac{R_{i}}{\sqrt{\pi \alpha'}} \, . 
\end{eqnarray*}
Dropping the volume factor the last expression can be rewritten as 
\begin{equation} \label{Tp}
T_{p}^{1l}= \lim_{u\to \infty} S(u, a^{*}(u)) = T_{p}^{tree}(1 + g\tilde Z_{p}) 
\end{equation}
where 
$$
T_{p}^{tree}= \frac{1}{g}(\sqrt{2\pi \alpha'})^{-13}(2\pi \sqrt{\alpha'})^{D} 
$$ is the tree level tension of a Dp-brane with $p=25-D$ and 
\begin{equation} \label{Tp'}
\tilde Z_{p} =  \frac{1}{(\sqrt{4\pi})^{26}} \int_{0}^{\infty}\frac{dt}{2\pi} 
\left( \frac{2\pi}{t}\right)^{(25-p)/2} e^{2t}f^{-24}(t)  \, . 
\end{equation}
Thus the equations (\ref{Tp}), (\ref{Tp'}) give the one-loop corrected tensions of Dp-branes. 
Again the normalization of this result seems to be quite meaningful. In particular for $D=0$ we recover (\ref{T25}).
Up to possible differences in normalization it agrees with 
the expected one-loop correction based on  on-shell  string theory considerations \cite{Fischler}. 
  
\section{Discussion}
In this paper we have computed the one loop correction to the
tachyonic potential in order to investigate its contribution to the
tree level tachyonic condensation process. This calculation consisted of
two steps: The first step, which was well defined and unambigous,
 was the calculation of a string amplitude
as a function of the modulus. The second step, more arbitrary and open
to question, was the determination of the modular measure and the use of formula 
(\ref{S2}) at the one-loop level. 
The recipe we have used for this part of the
problem, although simple and natural, is somewhat arbitrary and lacks
firm foundation. Clearly, more work is needed to put these results
on a sounder foundation.
                                                          
Another question that needs further investigation is the treatment
of divergences. We have chosen to avoid the divergences due to the
presence of tachyons by a suitable analytic continuation. This method
generates a complex tachyon potential, which is to be expected on
the grounds of vacuum instability. An alternative possibility is to
appeal to Fischler-Susskind mechanism \cite{FS}. Again, more work is
needed to clarify the situation. There is also the problem of
the divergence due to the dilaton when $D<3$, which we did not
treat in this paper.  

Our computations  lead to  corrections to the tachyon potential (\ref{poten}) and to brane tensions (\ref{Tp}), 
(\ref{Tp'})  
that look quite 
meaningful. In particular based on our considerations we may give the following  qualitative  argument on   
the nature of higher loop corrections to the process describing the reduction of $D25$ brane into a lower dimensional brane. 
If the picture with the boundary states discussed in section 2 is correct then the n-loop correction will have a factor 
of $(e^{-a}Z_{0}(u))^{n-1}$. Furthermore it looks plausible that similar to the one-loop case 
the corrections to the $a^{*}(u)$, i.e., to the value of $a$ extremizing the action, will be negligible as $u\to \infty$. 
In that case we will get a correction that up to a constant factor  coincides with the appropriate 
n-loop partition function with Dirichlet and Neumann boundary conditions. Thus it looks like in this situation 
nothing happens to the effective string coupling constant. The processes we considered  describe only some 
 descent relations between branes. Of course even if one starts with a single D0 brane there are relevant perturbations 
that  drive the system further to the bottom of the tachyon potential. As discussed in \cite{Kutasov1}, \cite{Gerasimov2}, 
\cite{Shenker} the constant perturbation, which one may always switch on as long as the world sheet has a boundary, 
will  on one hand keep the effective string coupling small and on the other hand 
will damp any open string amplitude. It is not clear to us that this apparent damping factor will dominate over 
any other possible relevant perturbation. There may be a growing factor in the boundary state similar to the $Z_{0}(u)$  
that will compensate $e^{-a}$.

Finally one  should also understand if normalizations of the one-loop corrected Dp-brane tensions are in accord with the on-shell 
string theory considerations. We leave these questions for a future investigation.

\begin{center} {\bf Acknowledgements}  \\
A. K. wants to acknowledge a useful discussion with Barton Zwiebach. 
 \end{center} 

\end{document}